\documentclass[prd,preprint]{revtex4}
\usepackage{color}
\usepackage{dcolumn}
\usepackage{amsmath}
\usepackage{bm}
\begin{document}

\title{Minimal Length Uncertainty Relation and gravitational quantum well}
\author{Fabian \surname{Brau}}
\email[E-mail: ]{F.Brau@cwi.nl} 
\affiliation{CWI, P.O. Box 94079, 1090 GB Amsterdam, The Netherlands}
\author{Fabien \surname{Buisseret}}
\email[E-mail: ]{fabien.buisseret@umh.ac.be} 
\affiliation{Groupe de Physique Nucl\'{e}aire Th\'{e}orique,
Universit\'{e} de Mons-Hainaut,
Acad\'{e}mie universitaire Wallonie-Bruxelles,
Place du Parc 20, BE-7000 Mons, Belgium}
\date{\today}

\begin{abstract}
The dynamics of a particle in a gravitational quantum well is studied in the context of nonrelativistic quantum mechanics with a particular deformation of a two-dimensional Heisenberg algebra. This deformation yields a new short-distance structure characterized by a finite minimal uncertainty in position measurements, a feature it shares with noncommutative theories. We show that an analytical solution can be found in perturbation and we compare our results to those published recently, where noncommutative geometry at the quantum mechanical level was considered. We find that the perturbations of the gravitational quantum well spectrum in these two approaches have different signatures. We also compare our modified energy spectrum to the results obtained with the GRANIT experiment, where the effects of the Earth's gravitational field on quantum states of ultra cold neutrons moving above a mirror are studied. This comparison leads to an upper bound on the minimal length scale induced by the deformed algebra we use. This upper bound is weaker than the one obtained in the context of the hydrogen atom but could still be useful if the deformation parameter of the Heisenberg algebra is not a universal constant but a quantity that depends on the energetic content of the system.
\end{abstract}



\maketitle

\section{Introduction}
\label{sec:intro}
The study of theories characterized by a minimal observable length is an active area in theoretical physics, not only because of their intrinsic interest, but also because their existence is suggested by string theory and quantum gravity~\cite{ncom,amati,magg93,qg,amel97,seib99}. These theories rely mainly on a modification of the canonical commutation relations between the position and momentum operators. We will consider in this paper a particular case of such a modification, which has been previously obtained in the context of perturbative string theory (see for example Ref.~\cite{amati}). As an illustration, in one dimension, and in units such as $\hbar=c=1$, it reads
\begin{equation}\label{rel1}
	\left[\hat{X},\hat{P}\right]=i(1+\beta \hat{P}^2).
\end{equation}
$\beta$ is a small parameter, assumed to be positive. If $\beta=0$, Eq.~(\ref{rel1}) clearly reduces to the ordinary Heisenberg algebra. Such a commutation relation leads to the following uncertainty relation 
\begin{equation}\label{uncer}
	\Delta \hat{X}\geq \frac{1}{2}\left(\frac{1}{\Delta \hat{P} }+\beta \Delta\hat{P}\right),
\end{equation}
 which implies the existence of a minimal length~\cite{kemp95}
\begin{equation}
	\Delta x_0=\sqrt{\beta}.
\end{equation}

This particular modification of the Heisenberg algebra, and its extension to higher dimensions, has been extensively studied recently, see for example Refs.~\cite{brou99,chan02,dadi03,ques03,hoss04,hoss04b,ques04,benc05,ques05,fity06,hoss06,noza06,stet06}.

Remark that it has also been argued that such a deformed Heisenberg algebra could also be used to describe, as an effective theory, non-pointlike particles: Hadrons, quasi-particles, collective excitations,\dots~\cite{kemp97b}. 

A recent experiment, called GRANIT, is devoted to the study of quantum states of neutrons in the Earth's gravitational field. Roughly speaking, in this experiment, ultra cold neutrons are freely moving in the gravitational field above a mirror. This particular setup gives rise to a so-called gravitational quantum well. As a consequence, the energy spectrum of the neutrons in the gravitational field's direction is quantized, and the probability of observing a particle at a given height will be maximum at the classical turning point $h_n=E_n/mg$, for each energy $E_n$. That is indeed what is observed. More details can be found in Refs.~\cite{nesv02,nesv03,nesv05}. This experiment gives an opportunity to make a confrontation between observation and various theoretical models concerning quantum effects in gravity, including eventual signatures of the existence of an intrinsic minimal length. 
\par A first study of the gravitational quantum well in a noncommutative geometry has been performed in Ref.~\cite{bert05}. It was based on the two dimensional commutation relations 
\begin{equation}\label{rel2}
	\left[x_1,x_2\right]=i\theta, \ \left[p_1,p_2\right]=i\eta,\ \left[x_j,p_k\right]=i\delta_{jk},
\end{equation}
and upper bounds on the parameters $\theta,\, \eta$ have been obtained by comparison with the experimental results of Ref.~\cite{nesv05}. Let us note that Eqs.~(\ref{rel2}) lead to the following uncertainty relations
\begin{equation}\label{uncer2}
\Delta x_1\, \Delta x_2\geq \frac{|\theta|}{2}, \ \Delta p_1\, \Delta  p_2\geq \frac{|\eta|}{2}, \ \Delta x_j\, \Delta p_j\geq \frac{1}{2}, 	
\end{equation}
also suggesting the existence of a minimal length accessible by measurement. A discussion about the distinctions and links between the minimal length uncertainty relation~(\ref{uncer}) and space uncertainty relations of the form~(\ref{uncer2}) can be found in Ref.~\cite{Yone}. 

Our goal is to study the deviations from the usual gravitational quantum well caused by a two dimensional analog of the modified Heisenberg algebra~(\ref{rel1}), instead of the relations~(\ref{rel2}) already used in Ref.~\cite{bert05}. Our paper is organized as follows. In Sec.~\ref{sec:theory}, we present the Schr\"odinger equation with a deformed Heisenberg algebra following Ref.~\cite{brau99}. We particularize it to the case of the gravitational quantum well, and obtain an analytic form for the energy spectrum in perturbation in Sec.~\ref{sec:theory2}. Then, we discuss the comparison between our results, those of Ref.~\cite{bert05}, and those of GRANIT in Sec.~\ref{sec:discu}. Finally, our conclusions are given in Sec.~\ref{sec:conclu}.

\section{The main equation for a general potential}
\label{sec:theory}

The method we use is essentially the same than the one developed in Ref.~\cite{brau99}. Nevertheless we recall in this section the main lines to make the paper self-contained. The modified Heisenberg algebra studied here is defined in $d$ dimensions by the following commutation relations~\cite{chan02,kemp97b}
\begin{subequations}
\begin{eqnarray}
\left[\hat{X}_i,\hat{P}_j\right] &=& i\left(\delta_{ij}+\beta \delta_{ij} \hat{P}^2 +\beta' \hat{P}_i \hat{P}_j\right), \\
\left[\hat{P}_i,\hat{P}_j\right] &=& 0,\\
\left[\hat{X}_i,\hat{X}_j\right] &=& i \frac{(2\beta-\beta')+(2\beta+\beta')\beta \hat{P}^2}{(1+\beta \hat{P}^2)} \hat{P}_{\left[i\right.}\hat{X}_{\left.j\right]},
\end{eqnarray}
\end{subequations}
where $\hat{P}^2 = \sum_{i=1}^{d} \hat{P}_i \hat{P}_i$ and where $\beta, \beta' > 0$ are considered as small quantities of the first order. Let us note that $\hat{P}_{\left[i\right.}\hat{X}_{\left.j\right]}=\hat{P}_i\hat{X}_j-\hat{P}_j\hat{X}_i$. In this paper, we only study the case $\beta'=2\beta$, which leaves the commutation relations between the operators $\hat{X}_i$ unchanged at the first order in $\beta$, i.e.
\begin{subequations}\label{eq1}
\begin{eqnarray}
\left[\hat{X}_i,\hat{P}_j\right] &=& i\left(\delta_{ij}+\beta \delta_{ij} \hat{P}^2 +2\beta \hat{P}_i \hat{P}_j\right), \\
\left[\hat{P}_i,\hat{P}_j\right] &=& 0\approx\left[\hat{X}_i,\hat{X}_j\right].
\end{eqnarray}
\end{subequations}
\par The commutation relations (\ref{eq1}) constitute the minimal extension of the Heisenberg algebra and are thus of special interest. In this case, the minimal length is given by \cite{kemp97b}
\begin{equation}\label{minl}
	\Delta x_0=\sqrt{(d+2)\beta}.
\end{equation}
\par To calculate a spectrum for a given potential, we must find a representation of the operators $\hat{X}_i$ and $\hat{P}_i$, involving position variables
$x_i$ and partial derivatives with respect to these position variables, which satisfies Eqs.~(\ref{eq1}), and then solve the corresponding Schr\"{o}dinger equation:
\begin{equation}
\label{eq2}
\left[ \frac{\hat{P}^2}{2m} + V\left(\vec{\hat{X}}\right)\right]\, \Psi(\vec{x}\,) = E\, \Psi(\vec{x}\,).
\end{equation}
It is straightforward to verify that the following representation fulfills the relations (\ref{eq1}), at the first order in $\beta$,
\begin{eqnarray}
\label{eq3}
\nonumber
\hat{X}_i\ \Psi(\vec{x}\,) &=& x_i \Psi(\vec{x}\,), \\
\hat{P}_i\, \Psi(\vec{x}\,) &=& p_i \left(1+\beta \vec{p}\,^2\right) \Psi(\vec{x}\,),  
\end{eqnarray}
with $p_i=-i\, \partial/\partial x_i$.
Neglecting terms of order $\beta^2$, the Schr\"{o}dinger equation (\ref{eq2}) takes the form
\begin{equation}
\label{eq4}
\left[ \frac{\vec{p}\,^2}{2m} +\frac{\beta}{m} \vec{p}\,^4 + V(\vec{x}\,)\right]\, \Psi(\vec{x}\,) = E\, \Psi(\vec{x}\,).
\end{equation}
This is the main equation from which the influence of a non vanishing $\beta$ can be studied. It also allows to compute upper bounds on this deformation parameter by comparison with experimental results. This equation is just the ordinary Schr\"{o}dinger equation with an additional term proportional to $\vec{p}\,^4$. As this correction is assumed to be small, we will compute its effects on the energy spectrum at the first order in perturbation.

\section{The main equation for a gravitational quantum well}
\label{sec:theory2}

Let us now consider the case of a particle of mass $m$, moving in a $z y$ plane, and subject to the Earth's gravitational field: $\vec{g}=-g\, \vec{e}_z$, where $g=9.80665$ m\, s$^{-2}$. In order to form a gravitational quantum well, a mirror is placed at $z=0$ and acts as an hardcore interaction. This corresponds to the experimental setup described in Refs.~\cite{nesv02,nesv03,nesv05}. It is reasonable to keep a constant value for $g$ because of the small size of the experiment. Taking into account the variation of the gravitational field would only introduce higher order corrections that can be neglected in this first order calculation. For the same reason, corrections coming from the fact that the mirror is not a perfect hardcore will not be included here. The potential which enters in Eq.~(\ref{eq4}) is then $V(\vec{x}\,)=V(z)$ with
\begin{eqnarray}
\label{eq4a}
V(z)&=& +\infty \quad \text{for} \quad z\le 0 \nonumber \\
    &=& mg z \quad \text{for} \quad z>0.
\end{eqnarray}
An infinite potential in $z=0$ is a very good description of the mirror, at least for the lowest eigenstates.

\subsection{The case $\beta=0$}

The solution of the Schr\"odinger equation in this context for $\beta=0$ is well known \cite[p. 101]{Flu}. We write
\begin{equation}
\label{eq4b}
\Psi(\vec{x}\,)=\psi_n(z)\psi(y).
\end{equation}
The wave function along the $z$ axis then reads
\begin{equation}
\label{eq4c}
\psi_n(z)=A_n\, {\rm Ai}(\bar{z}),\quad\text{with} \quad \bar{z}=\gamma z+\alpha_n \quad \text{and} \quad \gamma=(2m^2 g)^{1/3},
\end{equation}
where the function ${\rm Ai}(\bar{z})$ is the normalizable Airy function and where $\alpha_n$ are the zeros of this function. Their values can be found for example in Ref.~\cite[p. 478]{Abra}. 
The normalization factor $A_n$ is given by
\begin{equation}
\label{eq4d}
A_n=\left[ \frac{1}{\gamma}\int_{\alpha_n}^{+\infty} d\bar{z}\, {\rm Ai}^2(\bar{z})\right]^{-1/2}= \frac{\gamma^{1/2}}{|{\rm Ai}'(\alpha_n)|},
\end{equation}
where ${\rm Ai}'(x)$ is the derivative of the Airy function. The eigenvalues are related to the zeros of the Airy function as follows
\begin{equation}
\label{eq4e}
E_n^0=-\frac{m g}{\gamma}\, \alpha_n, 
\end{equation}
and consequently, the most probable heights where a particle can be detected are given by
\begin{equation}\label{eq4e2}
h_n^0=-\frac{\alpha_n}{\gamma}. 
\end{equation}
Along the $y$ axis, the particle is free and the wave function takes the form
\begin{equation}
\label{eq4f}
\psi(y)=\int_{-\infty}^{+\infty} dk\, g(k)\, e^{iky}, 
\end{equation}
where $g(k)$ determines the shape of the wave packet in momentum space.  

\subsection{The case $\beta>0$}

As $E_{n}^0$ denotes the eigenvalues for $\beta=0$, the energy spectrum to the first order in the deformation parameter $\beta$ is given by
\begin{equation}
\label{eq5}
E_{n}=E_{n}^0+\Delta E_{n}.
\end{equation}
It can be seen from Eq.~(\ref{eq4}) that the correction to the energy $\Delta E_{n}$ at the first order in $\beta$ reads
\begin{eqnarray}
\label{eq6}
\Delta E_{n}&=&\frac{\beta}{m}\langle \Psi(\vec{x}\, )|\vec{p}\,^4 |\Psi(\vec{x}\, ) \rangle, \nonumber \\
            &=&\frac{\beta}{m}\left[\langle \psi_n(z) |p_z^4 | \psi_n(z) \rangle + 2 \langle \psi_n(z) |p_z^2 | \psi_n(z) \rangle\langle \psi(y)
|p_y^2 | \psi(y) \rangle \right],
\end{eqnarray}
where a term proportional to $\langle \psi(y) |p_y^4 | \psi(y) \rangle$ has been omitted since it only leads to a global shift of the energy spectrum and has thus no interest. This last relation can be written as 
\begin{eqnarray}
\label{eq7}
\Delta E_{n}&=&\frac{\beta}{m}\left[4m^2\langle (E_n^0-V(z))^2   \rangle + 8m^2 E_c \langle  E_n^0-V(z)  \rangle\right], \nonumber \\ 
            &=&4\beta m\left[E_n^0 (E_n^0+2E_c)-2(E_n^0+E_c) \langle  V(z) \rangle+ \langle  V(z)^2  \rangle \right],            
\end{eqnarray}
where $E_c=m\langle \psi(y) |v_y^2|\psi(y)  \rangle / 2$ is the kinetic energy of the particle along the $y$ axis. The averages appearing in Eq.~(\ref{eq7}) are obviously computed with $\psi_n(z)$.
Since we consider here a power-law potential, $V(z)\sim z$, the virial theorem gives
\begin{equation}
\label{eq8}
\langle V(z) \rangle= \frac{2}{3}\, E_{n}^0,
\end{equation}
and the relation (\ref{eq7}) reduces to
\begin{eqnarray}
\label{eq9}
\Delta E_{n}&=&4\beta m\left[-\frac{E_n^0}{3} (E_n^0-2E_c)+ \langle  V(z)^2 \rangle \right], \nonumber \\
            &=&4\beta m\left[-\frac{E_n^0}{3} (E_n^0-2E_c)+(m g)^2 A_n^2 \int_0^{+\infty} dz\ z^2 \, {\rm Ai}^2(\gamma z+\alpha_n)\right].
\end{eqnarray}
The last integral in (\ref{eq9}) can be computed explicitly. We obtain
\begin{equation}
(m g)^2 A_n^2\int_0^{+\infty} dz\ z^2 \, {\rm Ai}^2(\gamma z+\alpha_n)=\frac{8}{15} (E_n^0)^2.
\end{equation}
The final result is then
\begin{equation}
\label{eq10}
\Delta E_{n}=\frac{4}{5}\beta m (E_n^0)^2\left(1+\frac{10 E_c}{3 E_n^0} \right).
\end{equation}

\section{Discussion of the results}\label{sec:discu}

\subsection{Comparison with noncommutative geometry}

Formula~(\ref{eq10}) involves the kinetic energy of the neutrons along the $y$ axis. The last term in the parenthesis of this formula is much larger than 1 (about 6 order of magnitude larger): $E^0_n$ has a value around few peV on Earth (see Eqs.~(\ref{eq4e}) and (\ref{engr})), and even for the nonrelativistic neutrons of Ref.~\cite{nesv05}, the kinetic energy is around $100$ neV. More precisely, the neutrons mean horizontal velocity was measured to be around 6.5 m\ s$^{-1}$ \cite{nesv05}. So, the kinetic energy is roughly equal to $E_c \simeq m\langle v_y \rangle^2 /2 \simeq\, 0.221\ \mu$eV (taking for the neutron mass the experimental value $m = 939.57$ MeV~\cite{pdg}). Clearly, we can thus use in a very good approximation 
\begin{equation}
\label{eq11}
\Delta E_{n}\cong \frac{8}{3}m\beta E_c\, E_n^0.
\end{equation}
This result can be compared to the energy shifts obtained with the noncommutative geometry~(\ref{rel2}). To the first order in the small parameters $\theta,\, \eta$, it is shown in Ref.~\cite{bert05} that these shifts, denoted $\Delta\tilde{
E}_n $, are given by
\begin{equation}\label{nc1}
\Delta \tilde{E}_n=\frac{\eta\langle v_y\rangle}{2}\left[\gamma^{-2}A^2_n\int^{+\infty}_{\alpha_n}ds\ s\
{\rm Ai}^2(s)-\frac{\alpha_n}{\gamma}\right].
\end{equation}
The integral appearing in Eq.~(\ref{nc1}) can be analytically computed and we obtain
\begin{equation}\label{nc2}
	\Delta \tilde{E}_n=\frac{\eta \langle v_y\rangle}{3mg}  E_n^0.
\end{equation}
\par A nonzero parameter $\beta$ leads, independently of its actual value, to a larger value for the heights since $h_n$~$=$~$h_n^0+\Delta h_n$, with $\Delta h_n=\Delta E_{n}/mg $ and $\Delta E_{n}$ given by Eq.~(\ref{eq11}). We recall indeed that $\beta$ was assumed to be positive. The signature of the modified Heisenberg algebra~(\ref{eq1}) would then be $\Delta h_n > 0$ and  
\begin{equation}\label{dh1}
\Delta h_n \propto \langle v_y\rangle^2 \left[\frac{3\pi}{2}(n-1/4)\right]^{2/3},	
\end{equation}
the second factor being a WKB approximation of the zeros of the Airy function \cite[p. 450]{Abra}. This effect is different of the one predicted by the relation~(\ref{nc2}), since in this case we should observe 
\begin{equation}\label{dh2}
\Delta \tilde{h}_n \propto \pm \langle v_y\rangle\, \left[\frac{3\pi}{2}(n-1/4)\right]^{2/3}. 	
\end{equation}
The $\pm$ factor in Eq.~(\ref{dh2}) arises from the fact that the sign of $\eta$ is a priori unspecified. So, if $\eta<0$, noncommutative geometry could cause a decrease of the heights instead of the increase predicted by our deformed algebra. Even if we suppose that $\eta>0$, the heights shifts~(\ref{dh1}) and (\ref{dh2}) can be distinguished in principle since their dependence on $\langle v_y\rangle$ is different. Another important difference between $\Delta h_n$ and $\Delta \tilde{h}_n$ concerns their dependence on the mass of the particle: $\Delta h_n \propto m^{4/3}$ whereas $\Delta \tilde{h}_n \propto m^{-5/3}$. 

\subsection{Comparison with GRANIT experiment}

We turn now our attention to the comparison between formula~(\ref{eq11}) and GRANIT results. Following Eqs.~(\ref{eq4e}) and (\ref{eq4e2}), the theoretical energies and heights for the two first eigenstates are
\begin{eqnarray}\label{engr}
	E^{0}_1&=&1.407\ {\rm peV}, \quad E^{0}_2=2.461\ {\rm peV},\nonumber \\
	h^{0}_1&=&13.7\  \mu{\rm m}, \quad\quad h^{0}_2=24.0\  \mu{\rm m}.
\end{eqnarray}
The experimental results concerning these states are~\cite{nesv05}
\begin{eqnarray}
	h^{{\rm exp}}_1&=&12.2\ \mu{\rm m}\pm1.8_{{\rm syst}}\pm0.7_{{\rm stat}}, \nonumber \\
	h^{{\rm exp}}_2&=&21.6\ \mu{\rm m}\pm2.2_{{\rm syst}}\pm0.7_{{\rm stat}}.
\end{eqnarray}
The theoretical values are contained in the error bars. The energy shifts due to eventual new physical mechanisms are thus bounded. They can not exceed
	\begin{eqnarray}\label{bound1}
	\Delta E^{{\rm exp}}_1&=&0.102\ {\rm peV},\nonumber \\
  \Delta E^{{\rm exp}}_2&=&0.051\ {\rm peV},
\end{eqnarray} 
since $\Delta E_{n}$ has been shown to be positive, see Eq.~(\ref{eq11}).
Consequently, we have to satisfy the constraint 
\begin{equation}
	\Delta E_{n}<\Delta E^{{\rm exp}}_n,
\end{equation}
with $\Delta E_n$ given by Eq.~(\ref{eq11}). Let us discuss the possible values of $\beta$. There are two possibilities, following the status of this parameter. 

The first possibility is that $\beta$ could be a new universal constant, or a function of the already known constants. Then, its value could be equally measured by independent experiments and a unique value, within error bars, would be found. It was shown in Ref.~\cite{brau99} how the spectrum of the hydrogen atom would be affected by the deformed Heisenberg algebra we use, and an upper bound $\Delta x_0< 10^{-2}$ fm was derived, or equivalently $\beta<2\, 10^{-5}$~fm$^2$ $\simeq 5\, 10^{-22}$ eV$^{-2}$. This estimation was based on a very precise (up to 1kHz) measurement of the radiation emitted during the transition 1S-2S of the hydrogen atom \cite{hydro}. With this upper bound, it is readily computed that  
\begin{equation}\label{bound2}
	\Delta E_1 \simeq \Delta E_2 \alt 10^{-19}\ {\rm peV}.
\end{equation}
If $\beta$ is a universal constant, the upper bounds~(\ref{bound2}) tell us that the effects of the existence of a minimal observable length are largely unobservable in the GRANIT experiment, since the maximal precision is $10^{-2}$ peV  \cite{nesv05}. 

The second possibility is that $\beta$ could vary from one system to another depending for example on the energetic content of the system (mass of the particles, strength of the interactions,\dots). If $\beta$ is such a quantity, the upper bound of Ref.~\cite{brau99} is no longer relevant for our study of neutrons in a gravitational quantum well (the mass of the particle and the interaction are different), and a new upper bound has to be determined from the experimental results. Equation~(\ref{eq11}), together with the relation~(\ref{minl}) for $d=2$ leads to   
\begin{equation}
\label{eq122}
\Delta x_0 < 2 \sqrt{\frac{3\Delta E_n^{\text{exp}}}{8 m E_n^0\, E_c}}.
\end{equation}
For $n=2$ we find $\Delta x_0< 0.012$ eV$^{-1}=2.41$ nm, or $\beta<1.46$ nm$^2$. The case $n=1$ does not lead to a better upper bound. This new upper bound for $\beta$ could be used to restrict the possible choices for an explicit expression of this parameter, in a similar way as the upper bound found in Ref.~\cite{brau99} was used in Ref.~\cite{brau03} to show that $\beta$ could not be identified to the Compton length of the particle as proposed in Ref.~\cite{sast00}: The effects of a minimal length $\Delta x_0$ on the energy spectrum would then be too large in the case of the hydrogen atom.

At last, remark that if $\Delta x_0$ is related to the size of the particle, as suggested in Ref.~\cite{kemp97b}, and discussed in the case of an electron in Ref.~\cite{brau99}, then $\Delta E_1 \simeq \Delta E_2 \alt 10^{-15}$ peV, since the size of the neutron is around 1 fm.

\section{Summary of the results}
\label{sec:conclu}

We have found in perturbation the energy spectrum of a gravitational quantum well with a one-parameter deformed Heisenberg algebra. This deformation implies in particular the existence of a minimal observable length, which is a feature it shares with usual noncommutative theories. We found that the energy shifts caused by this deformed algebra are positive with a linear dependence on the kinetic energy of the particles. This signature is different from the one coming from a previously studied noncommutative geometry \cite{bert05}. In this case indeed, the energy shifts can be either positive or negative, following the sign of the noncommutativity parameter, and they depend on the square root of the kinetic energy. The gravitational quantum well thus appears as an interesting physical system which allows, at least in principle, to distinguish between several approaches predicting different modifications of the Heisenberg algebra. 

By particularizing our results to the case of a neutron in the Earth's gravitational field, we are able to compare them to those of the GRANIT experiment. Our conclusion is twofold, following the status of the deformation parameter $\beta$. 
If $\beta$ is a universal constant, we can use an upper bound obtained previously by analysis of the hydrogen atom spectrum \cite{brau99}, and we find that the energy shifts due to a nonzero value of $\beta$ are around $10^{-19}$ peV. This is far beyond the experimental precision. 
However, if $\beta$ is a quantity that depends on the energetic content of the system (like the mass of the particle), we can derive a new upper bound from the GRANIT results. We conclude in that case that the minimal length scale associated to neutrons moving in a gravitational quantum well is smaller than few nanometers. This new upper bound could be used to constrain possible choices for an explicit expression of $\beta$.

\end{document}